\documentclass[11pt]{article}
\usepackage{cite}
\usepackage{mathrsfs}
\usepackage{amssymb,amsfonts,amsmath}
\usepackage{enumerate}
\usepackage{indentfirst}
\usepackage{latexsym,bm}
\usepackage[noend]{algpseudocode}
\usepackage{algorithmicx,algorithm}
\usepackage{algorithm}
\usepackage{makeidx}
\usepackage{fancybox}
\usepackage{color}
\usepackage{multicol}
\usepackage[latin1]{inputenc}
\usepackage{graphicx}
\usepackage{epstopdf}
\usepackage{pstricks,pst-node,pst-text,pst-3d}
\usepackage{tikz}
\newtheorem{thm}{Theorem}[section]

\newtheorem{lem}[thm]{Lemma}

\newenvironment{pf}{{\noindent \it \bf Proof:}}{{\hfill$\Box$}\\}

\begin{document}

\title{\bf Arc-disjoint in- and out-branchings rooted at the same vertex in compositions of digraphs}
\author{Gregory Gutin$^{1}$ and
Yuefang Sun$^{2}$\thanks{Corresponding author. This author was supported by NSFC No.
11401389.}\qquad\\
$^{1}$ Department of Computer Science\\ Royal Holloway, University of London\\ Egham, Surrey, TW20 0EX, UK\\ g.gutin@rhul.ac.uk\\
$^{2}$ Department of Mathematics, Shaoxing University\\ Zhejiang 312000, P. R. China, yuefangsun2013@163.com}

\date{}
\maketitle

\begin{abstract}
A digraph $D=(V, A)$ has a good pair at a vertex $r$ if $D$ has a pair of arc-disjoint in- and out-branchings rooted at $r$.
Let $T$ be a digraph with $t$ vertices $u_1,\dots , u_t$ and let $H_1,\dots H_t$ be digraphs such that $H_i$ has vertices $u_{i,j_i},\ 1\le j_i\le n_i.$ 
Then the composition $Q=T[H_1,\dots , H_t]$ is a digraph with vertex set $\{u_{i,j_i}\mid 1\le i\le t, 1\le j_i\le n_i\}$ and arc set $$A(Q)=\cup^t_{i=1}A(H_i)\cup \{u_{ij_i}u_{pq_p}\mid u_iu_p\in A(T), 1\le j_i\le n_i, 1\le q_p\le n_p\}.$$

When $T$ is arbitrary, we obtain the following result: every strong digraph composition $Q$ in which $n_i\ge 2$ for every $1\leq i\leq t$, has a good pair at every vertex of $Q.$ The condition of $n_i\ge 2$ in this result cannot be relaxed.
When $T$ is semicomplete, we characterize semicomplete compositions with a good pair, which generalizes the corresponding characterization by Bang-Jensen and Huang (J. Graph Theory, 1995) for quasi-transitive digraphs. As a result, we can decide in polynomial time whether a given semicomplete composition has a good pair rooted at a given vertex.
\vspace{0.3cm}\\
{\bf Keywords:} branching; semicomplete digraph; digraph composition.
\vspace{0.3cm}\\ {\bf AMS subject
classification (2010)}: 05C20, 05C70, 05C76, 05C85.

\end{abstract}

\section{Introduction}
We use a standard digraph terminology and notation as in \cite{Bang-Jensen-Gutin, Bang-Jensen-Gutin2}.
A digraph $D = (V, A)$ is {\em strongly connected} (or {\em
strong}) if there exists a path from $x$ to $y$ and a path from $y$
to $x$ in $D$ for every pair of distinct vertices $x, y$ of $D$.
An {\em out-tree} ({\em in-tree}, respectively) {\em rooted at a vertex $r$} is an orientation of a tree such that the in-degree (out-degree, respectively) of every vertex but $r$ equals one.
An {\em out-branching} ( {\em in-branching}, respectively) in a digraph $D$ is a spanning subgraph of $D$ which is out-tree (in-tree, respectively). It is well-known and easy to show  \cite{Bang-Jensen-Gutin, Bang-Jensen-Gutin2}
that a digraph has an out-branching (in-branching, respectively) rooted at $r$ if and only if $D$ has a unique initial strong connectivity component (terminal strong connectivity component, respectively) and $r$ belongs to this component.
Out-branchings and in-branchings when they exist can be found in linear-time using, say, depth-first search from the root.

Edmonds \cite{Edmonds} characterized digraphs with $k$
arc-disjoint out-branchings rooted at a specified vertex $r.$
Furthermore, there exists a polynomial algorithm for finding $k$
arc-disjoint out-branchings with a given root $r$ if they exist \cite{Bang-Jensen-Gutin}).
However, it is NP-complete to decide whether a digraph $D$ has a pair of arc-disjoint out-branching and in-branching rooted at $r,$ which was proved by Thomassen, see \cite{Bang-Jensen}.
Following \cite{Bang-Jensen-Huang2014} we will call such a pair a {\em good pair rooted at $r$}. Note that a good pair forms a strong spanning subgraph of $D$ and thus if $D$ has a good pair, then $D$ is strong.
The problem of the existence of a good pair was studied for tournaments and their generalizations, and characterizations (with proofs
implying polynomial-time algorithms for finding such a pair) were obtained in \cite{Bang-Jensen} for tournaments,  \cite{Bang-Jensen-Huang95}
for quasi-transitive digraphs and \cite{Bang-Jensen-Huang2014} for locally semicomplete digraphs. Also, Bang-Jensen and Huang \cite{Bang-Jensen-Huang95}  showed that if $r$ is adjacent to every vertex of $D$ (apart from itself) then $D$
has a good pair rooted at $r$.

In this paper, we study the existence of good pairs for digraph compositions.  Let $T$ be a digraph with $t$ vertices $u_1,\dots , u_t$ and let $H_1,\dots H_t$ be digraphs such that $H_i$ has vertices $u_{i,j_i},\ 1\le j_i\le n_i.$
Then the {\em composition} $Q=T[H_1,\dots , H_t]$ is a digraph with vertex set $\{u_{i,j_i}\mid 1\le i\le t, 1\le j_i\le n_i\}$ and arc set
$$A(Q)=\cup^t_{i=1}A(H_i)\cup \{u_{ij_i}u_{pq_p}\mid u_iu_p\in A(T), 1\le j_i\le n_i, 1\le q_p\le n_p\}.$$
Digraph compositions generalize some families of digraphs. In particular,   {\em semicomplete compositions}, which are compositions where $T$ is a semicomplete digraph,
generalize strong quasi-transitive digraphs as every strong quasi-transitive digraph is a strong semicomplete composition in which $H_i$ is either a one-vertex digraph or a non-strong quasi-transitive digraph.
To see that strong compositions form a significant generalization of strong quasi-transitive digraphs, observe that the Hamiltonian cycle problem is polynomial-time solvable for quasi-transitive digraphs \cite{gutinAJC10},
but NP-complete for strong compositions (see, e.g., \cite{Bang-Jensen-Gutin-Yeo}).
When $H_i$ is the same digraph $H$ for every $i\in [t]$, $Q$ is the lexicographic product of $T$ and $H$, see, e.g., \cite{Hammack}. While digraph compositions has been used since 1990s to study quasi-transitive digraphs and their generalizations, see, e.g., \cite{bangDM164,Bang-Jensen-Gutin,GSHC}, the study of digraph decompositions in their own right was initiated only recently in \cite{Sun-Gutin-Ai}.

In the next section, we obtain the following somewhat surprising result: every strong digraph composition $Q$ in which $n_i\ge 2$ for every $i\in [t]$, has a good pair rooted at every vertex of $Q.$ The condition of $n_i\ge 2$ in this result cannot be relaxed. Indeed, the characterization of quasi-transitive digraphs with  a good pair \cite{Bang-Jensen-Huang95} provides an infinite family of strong quasi-transitive digraphs which have no good pair rooted at some vertices. In Section \ref{sec:3}, we characterize semicomplete compositions with a good pair generalizing the corresponding result in \cite{Bang-Jensen-Huang95}. This allows us to decide in polynomial time whether a given semicomplete composition has a good pair rooted at a given vertex. In Section \ref{sec:4}, we discuss some open problems and a recent related result. 

Let $p$ and $q$ be integers. Then $[p..q]:=\{p,p+1,\dots ,q\}$ if $p\leq q$ and $\emptyset$, otherwise. In particular, if $p>0$,  $[p]$ will be a shorthand for $[1..p].$

\section{Compositions of digraphs: $T$ arbitrary}\label{sec:2}




A digraph $D=(V,A)$ has a {\em strong arc decomposition} if $A$
 has two disjoint sets $A_1$ and $A_2$ such that both $(V,A_1)$ and $(V,A_2)$ are strong.
Sun et al. \cite{Sun-Gutin-Ai} obtained sufficient conditions for a digraph composition to have a strong arc decomposition. In particular, they proved the following:

\begin{thm}\label{comp-good}
Let $T$ be a digraph with $t$ vertices  ($t\ge 2$) and let $H_1,\dots ,H_t$ be digraphs.
Then $Q=T[H_1,\dots ,H_t]$ has a strong arc decomposition if $T$ has a Hamiltonian cycle and one of the following conditions holds:
\begin{itemize}
\item $t$ is even and $n_i\ge 2$ for every $i=1,\dots ,t;$
\item $t$ is odd, $n_i\ge 2$ for every $i=1,\dots ,t$ and at least two distinct subgraphs $H_i$ have arcs;
\item  $t$ is odd and $n_i\ge 3$ for every $i=1,\dots ,t$ apart from one $i$ for which $n_i\ge 2$.
\end{itemize}
\end{thm}




\begin{lem}\label{lem1}
Let $Q=T[H_1,\dots ,H_t],$ where $t\ge 2.$ If $T$ has a Hamiltonian cycle and $H_1,\dots ,H_t$ are arbitrary digraphs, each with at least two vertices, then $Q$ has a good pair at any root $r$.
\end{lem}
\begin{pf} For the case that $t$ is even, by Theorem~\ref{comp-good}, $Q$ has has a pair of arc-disjoint strong spanning subgraphs $Q_1$ and $Q_2$. Observe that in $Q_1$ ($Q_2$, respectively), we can find an out-branching (in-branching, respectively) at $r$ (in polynomial time), as desired.

Now we assume that $t$ is odd. Without loss of generality, let $u_{1, 1}$ be the root. Let $T'_1$ be the path $u_{1, 1}u_{2, 1} \dots u_{t, 1} u_{1, 2} u_{2, 2} \dots u_{t, 2}$, and let $T'_2$ be the in-tree rooted at $u_{1, 1}$ with arc set $\{u_{i,2}u_{i+1,1}\mid 1\leq i\leq t-1\}\cup \{u_{i,1}u_{i+1,2}\mid 2\leq i\leq t-1\}\cup \{u_{t,1}u_{1,1}, u_{t,2}u_{1,1}\}$. By definition, $V(T'_1)=V(T'_2)=\{u_{i,j}\mid 1\leq i\leq t, 1\leq j\leq 2\}$. For any vertex $u_{i,j}$ with $1\leq i\leq t$ and $j\geq 3$, we add the arcs $u_{i-1,1}u_{i,j}$ and $u_{i,j}u_{i+1,1}$ to $T'_1$ and $T'_2$, respectively. Note that here $u_{0,1}=u_{t,1}$ and $u_{t+1,1}=u_{1,1}$. Observe that the resulting two subgraphs form a pair of out-branching and in-branching rooted at $u_{1,1}$, which are arc-disjoint.
\end{pf}

We will use the following decomposition of strong digraphs.
An {\em ear decomposition} of a digraph $D$ is a sequence
$\mathcal{P}=(P_0, P_1, P_2, \cdots, P_t)$, where $P_0$ is a cycle
or a vertex and each $P_i$ is a path, or a cycle with the following
properties:\\
$(a)$~$P_i$ and $P_j$ are arc-disjoint when $i\neq j$.\\
$(b)$~For each $i\in [0..t]$, let $D_i$ denote the digraph with
vertices $\bigcup_{j=0}^i{V(P_j)}$ and arcs
$\bigcup_{j=0}^i{A(P_j)}$. If $P_i$ is a cycle, then it has
precisely one vertex in common with $V(D_{i-1})$. Otherwise the end
vertices of $P_i$ are distinct vertices of $V(D_{i-1})$ and no other
vertex of $P_i$ belongs to $V(D_{i-1})$.\\
$(c)$~$\bigcup_{j=0}^t{A(P_j)}=A(D)$.

The following result is well-known, see, e.g.,
\cite{Bang-Jensen-Gutin}.
\begin{thm}\label{thm02}
Let $D$ be a digraph with at least two vertices. Then $D$ is strong
if and only if it has an ear decomposition. Furthermore, if $D$ is
strong, every cycle can be used as a starting cycle $P_0$ for an ear
decomposition of $D$, and there is a linear-time algorithm to find such
an ear decomposition.
\end{thm}

\begin{lem}\label{lem2}
Let $Q=T[\overline{K_2}, \dots, \overline{K_2}],$ where $|V(T)|=t\ge 2$ and $\overline{K_2}$ is the digraph with two vertices and no arcs.
If $T$ is strong, then $Q$ has a good pair at any root $r$.
\end{lem}
\begin{pf}
Without loss of generality, let $r=u_{1,1}$. Since $T$ is strong, $u_1$ belongs to some cycle $C$ in $T$.
By Theorem \ref{thm02}, $T$ has an ear decomposition $\mathcal{P}=(P_0, P_1, P_2,
\cdots, P_p)$, such that $P_0=C$ is the starting cycle. Let $G_i$ denote the
subgraph of $G$ with vertices $\bigcup_{j=0}^i{V(P_j)}$ and arcs
$\bigcup_{j=0}^i{A(P_j)}$.

We will prove the lemma by induction on $i\in \{0, 1, \dots, p\}$.
For the base step, by Lemma~\ref{lem1}, the subgraph $P_0[\overline{K_2}, \dots, \overline{K_2}]$ has a good pair rooted at $u_1$. For the inductive step, assume that $G_i[\overline{K_2}, \dots, \overline{K_2}]$ has a pair of arc-disjoint out-branching $T_1'$ and in-branching $T_2'$ rooted at $r$. Without loss of generality, let $P_{i+1}=u_su_{s+1}\dots u_{\ell}$. The following argument will be divided into two cases according to whether $P_{i+1}$ is a cycle.

\paragraph{Case 1: $P_{i+1}$ is a cycle.} In this case $u_s=u_{\ell}\in V(G_i)$. By Lemma~\ref{lem1}, in the subgraph $P_{i+1}[\overline{K_2}, \dots, \overline{K_2}]$, there is a pair of arc-disjoint out-branching $T_1''$ and in-branching $T_2''$ rooted at $u_{s,1}$. Let $T_1=T_1'\cup T_1''$ and $T_2=T_2'\cup T_2''$. Observe that $T_1$ is an out-branching and $T_2$ is an in-branching rooted at $r$ in $G_{i+1}[\overline{K_2}, \dots, \overline{K_2}].$
Since $P_{i+1}[\overline{K_2}, \dots, \overline{K_2}]$ and $G_i[\overline{K_2}, \dots, \overline{K_2}]$ are arc-disjoint,  $T_1$  and $T_2$ are arc-disjoint.

\paragraph{Case 2: $P_{i+1}$ is a path.} In this case, $u_s, u_{\ell}\in V(G_i)$ and $s\neq \ell$. We just consider the case that $\ell-s\geq 2$ since the remaining case is trivial (no need to change the current pair of out- and in-branchings). Let $T_1$ be the union of $T_1'$ and the two paths $u_{s,i}u_{s+1,i} \dots u_{\ell-1,i}$ where $1\leq i\leq 2$. 
Let $T_2$ be the union of $T_2'$ and the two paths $u_{s,1}u_{s+1,2}u_{s+2,1}u_{s+3,2}\dots u_{\ell,i}$ and $u_{s,2}u_{s+1,1}u_{s+2,2}u_{s+3,1}\dots u_{\ell,j},$ where $\{i,j\}=\{1,2\}.$
Observe that $T_1$ is an out-branching and $T_2$ is an in-branching rooted at $r$ in $G_{i+1}[\overline{K_2}, \dots, \overline{K_2}]$, moreover, they are arc-disjoint.

Thus, by induction, we are done.
\end{pf}

\begin{thm}\label{thma}
Let $Q=T[H_1,\dots ,H_t],$ where $t\ge 2.$ If $T$ is strong and $H_1,\dots ,H_t$ are arbitrary digraphs, each with at least two vertices, then $Q$ has a good pair at any root $r$. Furthermore, this pair can be found in polynomial time.
\end{thm}
\begin{pf}
Without loss of generality, let $r=u_{1,1}$. Let $Q'$ be the subgraph of $Q$ induced by the vertex set $\{u_{i,j}\mid 1\leq i\leq t, 1\leq j\leq 2\}$.
In $Q'$ delete arcs between vertices $u_{i,1}$ and $u_{i,2}$ for every $i\in [t].$
By Lemma \ref{lem2}, $Q'$ contains a pair of arc-disjoint out-tree $T_1'$ and in-tree $T_2'$ rooted at $r$.
By definition of out-tree, there is an arc $u_{p_i,q_i}u_{i,2}$ in $T_1'$ for every $i\in [t].$ For every $i\in [t]$ and $j\in [3..n_i],$ add $u_{p_i,q_i}u_{i,j}$ to $T_1'.$ This results in an out-branching $T_1.$
By definition of in-tree, there is an arc $u_{i,2}u_{a_i,b_i}$ in $T_2'$ for every $i\in [t].$ For every $i\in [t]$  and $j\in [3..n_i],$ add $u_{i,j}u_{a_i,b_i}$ to $T_2'.$ This results in an in-branching $T_2.$
Observe that $T_1$ and $T_2$ are arc-disjoint since $T'_1$ and $T'_2$ are arc-disjoint and the added arcs have heads and tails from $\{u_{i,j} \mid 1\leq i\leq t, 3\leq j\leq n_i\},$ respectively, in the arcs added to $T'_1$ and $T'_2,$ respectively.
Note that the proofs of Theorem \ref{comp-good}, Lemmas \ref{lem1} and \ref{lem2}, and this theorem are constructive and can be converted into polynomial-time algorithms. This fact and the polynomial-time algorithm
of Theorem \ref{thm02} imply that $T_1$ and $T_2$ can be constructed in polynomial time.
\end{pf}

\section{Compositions of digraphs: $T$ semicomplete}\label{sec:3}

We use $N^-(v)$ ($N^+(v)$, respectively) to denote the set of all in-neighbours (out-neighbours, respectively) of a vertex $v$ in a digraph $D$.


The next result was obtained by Bang-Jensen and Huang \cite{Bang-Jensen-Huang95}.

\begin{thm}\label{lem3}
Let $D$ be a strong digraph and $r$ a vertex of $D$ such that $V(D)=\{r\}\cup N^-(r)\cup N^+(r)$. There is a polynomial-time algorithm to decide whether $D$ has a good pair at $r$.
\end{thm}

For a path $P=x_1x_2\dots x_p$ and $1\le i\le j\le p$, let $P[x_i,x_j]:=x_ix_{i+1}\dots x_j.$
We now prove the following result on semicomplete compositions which generalizes a similar result for quasi-transitive digraphs by Bang-Jensen and Huang \cite{Bang-Jensen-Huang95}.

\begin{thm}\label{lem4}
A strong semicomplete composition $Q$ has a good pair rooted at $r$ if and only if $Q'=Q[\{r\}\cup N^-(r)\cup N^+(r)]$ has a good pair rooted at $r$.
\end{thm}
\begin{pf} Let $Q=T[H_1,\dots , H_t]$ and $A=V(Q)\setminus V(Q')$. Without loss of generality, assume that $r\in V(H_1)$. By definitions of a semicomplete composition and $Q'$, we have $A= V(H_1)\setminus \{r\}$.

Assume that $Q'$ has a good pair rooted at $r$, an out-branching $B'^+_r$ and an in-branching $B'^-_r$. Starting with $B'^+_r$,  we can construct an out-branching $B^+_r$ in $Q$ as follows. Let $v$ be a vertex such that  $vr\in B'^-_r$. Then add the arc $vu$ to $B'^+_r$ for each $u\in A.$ Similarly, starting with $B'^-_r$,  we could construct an in-branching $B^-_r$ in $Q$ as follows: for each $u\in A$, add the arc $uv'$ to $B'^-_r,$ where $rv'\in B'^+_r$. Observe that $B^+_r$ and $B^-_r$ are arc-disjoint, as desired.

Now we prove the other direction. Assume that $Q$ has a good pair, an out-branching $B^+_r$ and an in-branching $B^-_r$, rooted at $r$. If $B^+_r[V(Q')]$ and $B^-_r[V(Q')]$ are branchings, then we are done. Otherwise, we will obtain an in-branching (out-branching, respectively) from $B^-_r$ ($B^+_r,$ respectively) using the following procedure.

Choose a maximal path $P$ of $B^-_r$ to $r$, which contains a vertex $w\in A$, and assume that
$w$ is furthest from $r$ among vertices in $A\cap V(P)$. If $w$ is the first vertex of $P$, then delete it. Otherwise,
the previous vertex $u$ on $P$ has an arc $ur$ to $r$ (the arc $ur$ exists since $A\subseteq V(H_1)$), and we replace $P$ in $B^-_r$ by two paths: one is $P[p,u]r$, where $p$ is the first vertex of $P$, and the other is $P[w,r].$

Note that the in-degree $d^-(w)$ of $w$ has decreased by one. Thus, after $d^-(w)$ such replacements the in-degree of $w$ becomes equal to zero, i.e., $w$ is the first vertex on its maximal path $Q$ to $r$ and therefore $w$ will be deleted when we consider $Q.$ This means that after a finite number of replacements, we will delete all vertices of $A$ in $B^-_r$ and obtain an in-branching $B_r'^-$ of $Q'$ rooted at $r.$ Similarly, we can construct an out-branching $B'^+_r$ of $Q'.$ Note that to build $B_r'^-$ we add only arcs to $r$ and to build $B'^+_r$ we add only arcs from $r.$ This fact and the fact that $B^-_r$ and $B^+_r$ are arc-disjoint, imply that $B'^-_r$ and $B'^+_r$ are arc-disjoint, too.

\end{pf}

By Theorems \ref{lem3} and \ref{lem4},  we immediately have the following:

\begin{thm}\label{thmb}
Given a semicomplete composition and a vertex $r$, we can decide in polynomial time whether $D$ has a good pair rooted at $r.$
\end{thm}


\section{Open Problems and Related Results}\label{sec:4}

Theorem \ref{lem4} provides a characterization for the following problem for semicomplete compositions: given a digraph $D$ and a vertex $r\in V(D)$ decide whether $D$ has a good pair rooted at $r.$ The theorem generalizes a similar characterization by Bang-Jensen and Huang \cite{Bang-Jensen-Huang95} for quasi-transitive digraphs. Strong semicomplete compositions is not the only class of digraphs generalizing strong quasi-transitive digraphs. Other such classes have been studied such as $k$-quasi-transitive digraphs \cite{GSHC} and it would be interesting to see whether a characterization for the problem (or, at least non-trivial sufficient conditions) on $k$-quasi-transitive digraphs can be obtained. As we mentioned above,  Bang-Jensen and Huang \cite{Bang-Jensen-Huang2014} obtained a characterization for the problem on locally semicomplete digraphs. It would be interesting to see whether a characterization for the problem on in-locally semicomplete digraphs \cite{bang2018,Bang-Jensen-Gutin} can be obtained.

An out-branching and in-branching $B^+_r$ and $B^-_r$ are called $k$-{\em distinct} if $B^+_r$ has at least $k$ arcs not present in $B^-_r.$ The problem of deciding whether a digraph $D$ has a $k$-distinct pair of out- and in-branchings is NP-complete since it generalizes the good pair problem ($k=|V(D)|-1$).  Bang-Jensen and Yeo \cite{bangDAM156} asked whether the $k$-distinct problem is fixed-parameter tractable when parameterized by $k$, i.e., whether there is an $O(f(k)|V(D)|^{O(1)})$-time algorithm for solving the problem, where $f(k)$ is an arbitrary computable function in $k$ only. Gutin, Reidl and Wahlstr{\"o}m \cite{gutinJCSS95}  answered this open question in affirmative by designing an $O(2^{O(k\log^2 k)}|V(D)|^{O(1)})$-time algorithm for solving the problem.



\end{document}